\documentclass[12pt]{article}
\usepackage{amssymb}
\def\be{\begin{equation}}
\def\ee{\end{equation}}
\def\bea{\begin{eqnarray}}
\def\eea{\end{eqnarray}}
\def\p{\partial}

\def\pl#1{{\sl Phys.~Lett.~\bf B#1}}
\def\pr#1{{\sl Phys.~Rev.~\bf D#1}}
\def\prl#1{{\sl Phys.~Rev. Lett.~\bf #1}}

\def\cqg#1{{\sl Class.~Quant.~Grav.~\bf #1}}

\catcode`\@=11
\def\@citex[#1]#2{%
\if@filesw \immediate \write \@auxout {\string \citation {#2}}\fi
\@tempcntb\m@ne \let\@h@ld\relax \def\@citea{}%
\@cite{%
  \@for \@citeb:=#2\do {%
    \@ifundefined {b@\@citeb}%
      {\@h@ld\@citea\@tempcntb\m@ne{\bf ?}%
      \@warning {Citation `\@citeb ' on page \thepage \space undefined}}%
      {\@tempcnta\@tempcntb \advance\@tempcnta\@ne%
      \@tempcntb\number\csname b@\@citeb \endcsname \relax%
      \ifnum\@tempcnta=\@tempcntb %Number follows previous--hold on to it
        \ifx\@h@ld\relax%
          \edef \@h@ld{\@citea\csname b@\@citeb\endcsname}%
        \else%
          \edef\@h@ld{\ifmmode{-}\else--\fi\csname b@\@citeb\endcsname}%
        \fi%
      \else%   %  non-successor--dump what's held and do this one
        \@h@ld\@citea\csname b@\@citeb \endcsname%
        \let\@h@ld\relax%
      \fi}%
    \def\@citea{,\penalty\@highpenalty\,}%
  }\@h@ld
}{#1}}

\def\@citeb#1#2{{[#1]\if@tempswa , #2\fi}}
\def\@citeu#1#2{{$^{#1}$\if@tempswa , #2\fi }}
\def\@citep#1#2{{#1\if@tempswa , #2\fi}}

\def\bcites{         % cite with []'s
        \catcode`\@=11
        \let\@cite=\@citeb
        \catcode`\@=12
}

\def\upcites{         % cite with exponents
        \catcode`\@=11
        \let\@cite=\@citeu
        \catcode`\@=12
}

\def\plaincites{      % cite without brackets
        \catcode`\@=11
        \let\@cite=\@citep
        \catcode`\@=12
}

\newcount\hour
\newcount\minute
\newtoks\amorpm
\hour=\time\divide\hour by 60
\minute=\time{\multiply\hour by 60 \global\advance\minute by-\hour}
\edef\standardtime{{\ifnum\hour<12 \global\amorpm={am}%
        \else\global\amorpm={pm}\advance\hour by-12 \fi
        \ifnum\hour=0 \hour=12 \fi
        \number\hour:\ifnum\minute<10 0\fi\number\minute\the\amorpm}}
\edef\militarytime{\number\hour:\ifnum\minute<10 0\fi\number\minute}

\def\draftlabel#1{{\@bsphack\if@filesw {\let\thepage\relax
   \xdef\@gtempa{\write\@auxout{\string
      \newlabel{#1}{{\@currentlabel}{\thepage}}}}}\@gtempa
   \if@nobreak \ifvmode\nobreak\fi\fi\fi\@esphack}
        \gdef\@eqnlabel{#1}}
\def\@eqnlabel{}
\def\@vacuum{}
\def\marginnote#1{}
\def\draftmarginnote#1{\marginpar{\raggedright\scriptsize\tt#1}}
\def\draft{
        \pagestyle{plain}
        \overfullrule=2pt
        \oddsidemargin -.5truein
        \def\@oddhead{\sl \phantom{\today\quad\militarytime} \hfil
        \smash{\Large\sl DRAFT} \hfil \today\quad\militarytime}
        \let\@evenhead\@oddhead
        \let\label=\draftlabel
        \let\marginnote=\draftmarginnote
        \def\ps@empty{\let\@mkboth\@gobbletwo
        \def\@oddfoot{\hfil \smash{\Large\sl DRAFT} \hfil}
        \let\@evenfoot\@oddhead}
        \def\@eqnnum{(\theequation)\rlap{\kern\marginparsep\tt\@eqnlabel}%
        \global\let\@eqnlabel\@vacuum}  }
\begin{document}

\hfill UTHET-03-0701

%\hfill {\tt hep-th/yymmxxx} 
\vspace{-0.2cm}

\begin{center}
\Large
{\bf Asymptotic form of quasi-normal modes of large AdS black holes}
\normalsize

\vspace{0.8cm}
{\bf
Suphot Musiri}\footnote{email: suphot@swu.ac.th} \\
Department of Physics, \\
Srinakharinwiroth University, Bangkok, \\
Thailand.
\vspace{.5cm}

{\bf George Siopsis}\footnote{email: gsiopsis@utk.edu}\\ Department of Physics
and Astronomy, \\
The University of Tennessee, Knoxville, \\
TN 37996 - 1200, USA.

\end{center}

\vspace{0.8cm}
\large
\centerline{\bf Abstract}
\normalsize
\vspace{.5cm}

We discuss a method of calculating analytically
the asymptotic form of quasi-normal frequencies for large AdS black holes in
five dimensions.
In this case, the wave equation reduces to a Heun equation.
We show that the Heun equation may be approximated by a Hypergeometric equation
at large frequencies. Thus we obtain the asymptotic form of quasi-normal
frequencies in agreement with numerical results. We also present a simple
monodromy argument that leads to the same results.
We include a comparison with the three-dimensional case in which exact expressions are derived.

\newpage
%\section{Introduction}

Quasi-normal modes play an important role in the study of black holes and have
long been the subject of extensive research primarily in the case of
asymptotically flat space-times (for a review see~\cite{bibf1,bibf2}).
More recently, the asymptotic form of quasi-normal frequencies was shown to be
related to the Barbero-Immirzi parameter of Loop Quantum Gravity (see~\cite{bibf3} and references therein).
Quasi-normal modes form a discrete spectrum of complex frequencies whose imaginary part determines the decay rate of the fluctuation.
They are obtained as solutions to a wave equation subject to the conditions that the
flux be ingoing at the horizon and outgoing at asymptotic infinity.

There has been an extensive investigation of quasi-normal modes in the case
of asymptotically AdS space-times~\cite{bibq1,bibq2,bibq3,bibq4,bibq5,bibq6,bibq7,bibq8,bibq9,bibq10,bibq11,bibq12,bibq13,bibw1,bibw2,bibw3,bibw4}.
in conjunction with the AdS/CFT correspondence.
In this case the quasi-normal modes should correspond to perturbations of
the dual CFT.
The potential does not vanish in spatial infinity, so instead of demanding
outgoing flux there, as in asymptotically flat space-times, the wavefunction ought to vanish at infinity.
Exact results have been obtained in three space-time dimensions~\cite{bibq7,bibq13}.
In higher dimensions the wave equation cannot in general be solved analytically
and only numerical results on the quasi-normal frequencies have been obtained
in four, five and seven dimensions~\cite{bibq2,bibq14,bibr1,bibr2}.

In ref.~\cite{bibus}, we discussed an analytic method of calculating
the quasi-normal modes of
a large AdS black hole. The method was based on a
perturbative expansion
of the wave equation in the dimensionless parameter $\omega / T_H$, where $\omega$
is the frequency of the mode and $T_H$ is the (high) Hawking temperature of the black hole.
This is a non-trivial expansion, for the dependence of the wavefunction on $\omega / T_H$ changes as one moves from the asymptotic boundary of AdS space to the horizon
of the black hole.
The zeroth-order approximation was chosen to be an appropriate Hypergeometric equation so that higher-order corrections were indeed of higher order in $\omega /T_H$.
The first-order correction was also calculated.
We thus obtained an approximation to the low-lying quasi-normal frequencies.
We showed that our results were in agreement with numerical results~\cite{bibq2,bibq14} in five dimensions where the wave equation reduces to a Heun equation~\cite{bibq15}.

Here we discuss an approximation to the wave equation which is valid in the
high frequency regime instead. In five dimensions we show that the Heun equation
reduces to a Hypergeometric equation, as in the low frequency regime~\cite{bibus}.
We obtain an analytical expression for the asymptotic form of quasi-normal frequencies in agreement
with numerical results~\cite{bibq2,bibq14}.
These expressions may also be easily obtained by considering the monodromies
around the singularities of the wave equation. These singularities lie in the
unphysical region. In three dimensions, they are located at the horizon $r=r_h$,
where $r_h$ is the radius of the horizon, and at the black hole singularity,
$r=0$. In higher dimensions, it is necessary to analytically continue $r$ into
the complex plane. The singularities lie on the circle $|r| = r_h$. The
situation is similar to the case of asymptotically flat space where an
analytic continuation of $r$ yielded the asymptotic form of quasi-normal
frequencies~\cite{bibmo1,bibmo2}. It is
curious that unphysical singularities determine the behavior of quasi-normal
modes.

%\section{A large AdS black hole}

The metric of a $d$-dimensional AdS black hole may be written as
$$
ds^2 = -f(r)\, dt^2 + \frac{dr^2}{ f(r)} +r^2 d \Omega_{d-2}^2
$$
\be f(r) = \frac{r^2}{R^2} +1 - \frac{\omega_{d-1} M}{r^{d-3}}\ee
where $R$ is the AdS radius and $M$ is the mass of the black hole.
For a large black hole, the metric simplifies to
$$
ds^2 = -\hat f(r)\,  dt^2 + \frac{dr^2}{\hat f(r)} +r^2 ds^2 (\mathbb{E}^{d-2})
$$
\be \hat f(r) = \frac{r^2}{R^2} - \frac{\omega_{d-1} M}{r^{d-3}}\ee
The Hawking temperature is
\be
T_H = \frac{d-1}{4\pi}\; \frac{r_h}{R^2}
\ee
where $r_h$ is the radius of the horizon,
\be
r_h = R\; \left[ \frac{\omega_{d-1} M}{R^{d-3}} \right] ^{1/(d-1)}
\ee
The scalar wave equation is
\be\label{eq5}
\frac{1}{\sqrt{g}} \p_A \sqrt{g}g^{AB}\p _B \Phi = m^2 \Phi \ee
We are interested in solving this equation in the massless case ($m=0$) for
a wave which is ingoing at the horizon and vanishes at infinity. These boundary
conditions yield a discrete set of complex frequencies (quasi-normal modes).

%\section{Three dimensions}

We start with a review of the three-dimensional case where the assumption of a
large black hole is redundant and the wave equation may be solved exactly~\cite{bibq7,bibq13}.
Indeed, in three dimensions
$(d=3)$, the metric reads
\be
ds^2 = \frac{1}{R^2}\; \left( r^2 -r_h^2 \right) dt^2 +\frac{R^2\; dr^2}{ \left( r^2 - r_h^2 \right) }+ r^2 dx^2
\ee
independently of the size of the black hole.
The wave equation (with $m=0$) is
\be
\frac{1}{R^2\; r}\p_r \left( r^3 \left( 1- \frac{r_h^2}{r^2}\right) \p_r \Phi\right) -\frac{R^2}{r^2 - r_h^2 }\p_t^2 \Phi + \frac{1}{r^2}\p_x^2 \Phi =0 \nonumber
\ee
One normally solves this equation in the physical interval $r\in [r_h,\infty)$.
Instead, we shall solve it in the interval $0\le r\le r_h$ (inside the horizon).
The solution may be written as
\be
\Phi = e^{i(\omega t-px) }\Psi (y) ,\ \ \ \ \ y = \frac{r^2}{r_h^2}
\ee
where $\Psi$ satisfies
\be
\left( y (1-y) \Psi' \right)'
+ \left( \frac{\hat\omega^2}{1- y} +\frac{\hat p^2}{ y} \right)\Psi =0
\ee
in the interval $0<y<1$, and we have introduced the dimensionless variables
\be
\hat\omega = \frac{\omega R^2}{2r_h} = \frac{\omega}{4\pi T_H},\ \ \ \hat p^2 = \frac{pR}{2r_h} = \frac{p}{4\pi R T_H}
\ee
Two independent solutions are obtained by examining the behavior near the horizon
($y\to 1$),
\be\label{eq11} \Psi_\pm \sim (1-y)^{\pm i\hat\omega}\ee
where $\Psi_+$ is outgoing and $\Psi_-$ is ingoing. A different set of linearly
independent solutions is obtained by studying the behavior at the black hole
singularity ($y\to 0$).
We obtain
\be \Psi\sim y^{\pm i\hat p}
\ee
For quasi-normal modes, we demand that $\Psi$ be purely ingoing at the horizon
($\Psi \sim \Psi_-$ as $y\to 1$). By writing
\be \Psi (y) = y^{\pm i\hat p} (1-y)^{-i\hat\omega} F(y)\ee
we deduce
\be y(1-y) F'' + \{ 1\pm 2i\hat p -(2-2i(\hat\omega \mp\hat p)y \}\, F'
+ (\hat\omega \mp \hat p)(\hat\omega \mp\hat p +i)\, F =0 \ee
whose solution is the
Hypergeometric function
\be\label{eqhy} F (y) = {}_2F_1 (1-i(\hat\omega \mp \hat p), -i(\hat\omega \mp \hat p); 1\pm
2i\hat p; y)\ee
In general, near the horizon ($y\to 1$), this solution is a mixture of ingoing and outgoing waves.
It also blows up at infinity. To obtain the desired behavior for a quasi-normal
mode at $y=1$ and $y\to\infty$, we demand that $F(y)$ be a Polynomial.
This condition implies
\be\label{eqwme2} \hat\omega = \pm \hat p  -in\quad,\quad n=1,2,\dots\ee
a discrete set of complex frequencies with negative imaginary part, as expected~\cite{bibq2}. Notice that we obtained two sets of frequencies, with opposite real parts.
Then  eq.~(\ref{eqhy}) implies that
\be F(y) = {}_2F_1 (1-n,-n; 1\pm 2i\hat p; y)\ee
which is a Polynomial of order $n-1$.
It is therefore a constant at $y=1$, as desired and behaves as $F(y)\sim y^{n-1}
\sim y^{i(\hat\omega \mp \hat p)-1}$ as $y\to\infty$. Therefore,
$\Psi\sim y^{-1}$ as $y\to\infty$, as expected.

The above quasi-normal frequencies may also be deduced from a simple monodromy
argument. Let $\mathcal{M}(y_0)$ be the monodromy around the singular point $y=y_0$ computed along a small circle centered at $y=y_0$ running counterclockwise.
For $y=1$, we obtain
\be\mathcal{M} (1) = e^{2\pi\hat\omega}\ee
whereas for $y=0$, we have
\be\mathcal{M} (0) = e^{\mp 2\pi \hat p}\ee
Since the function vanishes at infinity, the two contours around the two singular points $y=0,1$ may be deformed into each other without encountering any
singularities. This implies that
\be\label{eqmo1}\mathcal{M} (1) \, \mathcal{M} (0) = 1\ee
hence $e^{2\pi (\hat\omega \mp \hat p)} = e^{2\pi i n}$ ($n\in\mathbb{Z}$), which leads to the same set of quasi-normal frequencies as before, if we demand
Im$\hat\omega < 0$.
Notice that the monodromy argument is much simpler here than in the case of
an asymptotically flat space-time~\cite{bibmo2}.
This is because of a simpler boundary condition at infinity ($y\to \infty$).

%\section{A large AdS black hole}

%\section{Five dimensions}

In five dimensions ($d=5$), the wave equation~(\ref{eq5}) with
$m=0$ reads
\be
\frac{1}{r^3}\p_r (r^5\, \hat f(r)\, \p_r \Phi) -\frac{R^4}{ r^2\,\hat f(r) }\p_{t}^2\Phi - \frac{R^2}{r^2}\; \vec\nabla^2\Phi = 0
\ee
where
\be \hat f(r) = 1- \frac{r_h^4}{r^4}
\ee
The solution may be written as
\be 
\Phi = e^{i(\omega t - \vec p\cdot \vec x)} \Psi (r)
\ee
Upon changing the coordinate $r$ to $y$,
\be
y = \frac{r^2}{r_h^2} 
\ee
the wave equation becomes
\be\label{eq24}
(y^2-1)\left( y(y^2-1) \Psi' \right)' + \left(\frac{\hat\omega^2}{4}\, y^2 - \frac{\hat p^2}{4}\, (y^2-1)\right)\Psi = 0
\ee
where we have introduced the dimensionless variables
\be
\hat\omega = \frac{\omega R^2}{r_h} = \frac{\omega}{\pi T_H}, \ \ \ \ \
\hat p = \frac{|\vec p|R}{r_h} = \frac{|\vec p|}{\pi R T_H}
\ee
Two independent solutions are obtained by examining the behavior near the
horizon ($y\to 1$),
%Near the horizon we obtain in- and out-going waves
\be\label{eqn1} \Psi_\pm \sim (y-1)^{\pm i\hat\omega/4}\ee
where $\Psi_+$ is outgoing and $\Psi_-$ is ingoing.
A different set of linearly independent solutions is obtained by studying the
behavior at large $r$
($y\to \infty$). We obtain
\be\label{eq27} \Psi\sim y^{h_\pm} \ \ , \ \ h_\pm = 0,-2\ee
so one of
the solutions contains logarithms. For quasi-normal modes, we are interested
in the analytic solution which behaves as $\Psi\sim y^{-2}$ as $y\to\infty$.
By considering the other (unphysical) singularity at $y=-1$, we obtain
another set of linearly independent wavefunctions behaving as
\be \Psi \sim (y+1)^{\pm \hat\omega /4} \ee
near $y=-1$. Following the discussion in the three-dimensional case,
we shall isolate the behavior at the two singularities $y=\pm 1$ and
write the wavefunction as
\be\label{eq25}
\Psi (y) = (y-1)^{-i\hat\omega/4} (y+1)^{\pm\hat\omega/4} F(y)
\ee
The two sets of modes have the same imaginary parts, but opposite
real parts, as we shall see (similarly to the $d=3$ case (eq.~(\ref{eqwme2}))).

It is easily deduced from eqs.~(\ref{eq24}) and (\ref{eq25}) that
the function $F(y)$ satisfies the Heun equation
\bea
y(y^2-1) F'' + \left\{ \left( 3- \frac{i\pm 1}{2}\, \hat\omega \right) y^2 - \frac{i \pm 1}{2}\, \hat\omega y -1 \right\} F' & & \nonumber \\
+ \left\{ \frac{\hat\omega}{2}\left( \pm \frac{i\hat\omega}{4} \mp 1-i\right) y - (i\mp 1)\frac{\hat\omega}{4} - \frac{\hat p^2}{4} \right\}\; F &=& 0 \label{w1}
\eea
We wish to solve this equation in a region in the complex $y$-plane
containing $|y|\ge 1$, which includes the
physical regime $r> r_h$.
For large $\hat\omega$, the constant terms in the respective Polynomial coefficients of $F'$ and $F$ in~(\ref{w1}) are small compared with the other terms, so they may be dropped.
Eq.~(\ref{w1}) may then be approximated by the Hypergeometric equation
\be
(y^2-1) F'' + \left\{ \left( 3- \frac{i\pm 1}{2}\, \hat\omega \right) y - \frac{i \pm 1}{2}\, \hat\omega \right\} F'
+ \frac{\hat\omega}{2}\left( \pm \frac{i\hat\omega}{4} \mp 1-i\right)\; F =0 \label{w2}
\ee
in the asymptotic limit of large frequencies $\hat\omega$.
The analytic solution of~(\ref{w2}) is the Hypergeometric function
\be\label{eq32} F_0(x) = {}_2F_1 ( a_+, a_-; c; (y+1)/2)\ee
where
\be a_\pm = 1-{\textstyle{\frac{i \pm 1}{4}}}\,\hat\omega\pm 1
\quad,\quad c = {\textstyle{\frac{3}{2}}}\pm {\textstyle{\frac{1}{2}}}\,\hat\omega\ee
For proper behavior at $y\to\infty$, we demand that $F$ be a {\em Polynomial}.
This requires
\be\label{eqqnf} a_+ = -n \ \ , \ \ n = 1,2,\dots\ee
Then $F$ is a Polynomial of order $n$, so at infinity it behaves as $F\sim y^n
\sim y^{-a_+}$.
The behavior of $\Psi$ is then deduced from (\ref{eq25}) to be
\be \Psi \sim y^{-i\hat\omega/4} y^{\pm\hat\omega/4} y^{-a_+} \sim y^{-2} \ee
as expected.
The quasi-normal frequencies from eq.~(\ref{eqqnf}) are then easily found to
be given by
\be\label{eqqnf5} \hat\omega = 2n(\pm 1-i) \ee
in agreement with numerical results~\cite{bibq14}.

These frequencies may also be obtained by a monodromy argument similar to the
one in $d=3$. If the function has no singularities other than $y=\pm 1$,
the contour around $y=+1$ may be unobstructedly deformed into the contour
around $y=-1$. Hence
\be\label{eqmo2} \mathcal{M} (1) \mathcal{M} (-1) = 1\ee
Since $\mathcal{M} (1) = e^{\pi \hat\omega /2}$, $\mathcal{M} (-1) = e^{\mp i\pi \hat\omega /2}$, the expression~(\ref{eqqnf5}) is easily deduced (also using
Im$\hat\omega < 0$).

The above approximation~(\ref{eq32}) to the exact wavefunction satisfying the
Heun equation~(\ref{w2}) may be used as the basis for a systematic calculation
of corrections to the asymptotic form~(\ref{eqqnf5}) of quasi-normal frequencies.
To this end,
let us write the Heun equation~(\ref{w2}) as
\be\label{w3} (\mathcal{H}_0 +\mathcal{H}_1) F = 0\ee
where
\be \mathcal{H}_0  = (1-y^2)\frac{d^2}{dy^2} +
\left\{ \frac{i\pm 1}{2}\, \hat\omega
- \left( 3- \frac{i\mp 1}{2}\, \hat\omega \right) y \right\}
\frac{d}{dy}
- \frac{\hat\omega}{2}\left( \pm \frac{i\hat\omega}{4} \mp 1-i\right)
\ee
and
\be\mathcal{H}_1 = \frac{1}{y} \left(\frac{d}{dy} + (i\mp 1)\frac{\hat\omega}{4} + \frac{\hat p^2}{4}
\right)\ee
The zeroth-order equation,
\be \mathcal{H}_0 F_0 = 0\ee
is the approximation we discussed above (the Hypergeometric eq.~(\ref{w2})).
By treating $\mathcal{H}_1$ as a perturbation, we may
expand the wavefunction,
\be F = F_0 + F_1 + \dots\ee
and solve eq.~(\ref{w3}) perturbatively. Corrections to the quasi-normal
frequencies~(\ref{eqqnf5}) may then be obtained once an explicit expression
for the corrections to $F_0$ have been calculated.
Details of this calculation will appear elsewhere~\cite{bibelse}.

In higher dimensions, the wave equation possesses more than two singularities on
the circle $|r| = r_h$ in the complex $r$-plane. Thus, a simple monodromy
argument such as the one discussed above in three and five dimensions~(eqs.~(\ref{eqmo1}) and (\ref{eqmo2}), respectively) does not appear to be applicable.
Work in this direction is in progress.

\section*{Acknowledgments}

G.~S.~is supported by the US Department of Energy under grant
DE-FG05-91ER40627.

%\newpage


\begin{thebibliography}{99}

\bibitem{bibf1} K~D.~Kokkotas and B.~G.~Schmidt, {\sl Living Rev.~Rel.~\bf 2}
(1999) 2; {\tt gr-qc/9909058}.
\bibitem{bibf2} H.~P.~Nollert, \cqg{16} (2000) R159.
\bibitem{bibf3} J.~Baez, in {\em Matters of gravity}, p.~12, ed.~J.~Pullin;
{\tt gr-qc/0303027}.
\bibitem{bibq1} J.~S.~F.~Chan and R.~B.~Mann, \pr{55} (1997) 7546; {\tt gr-qc/9612026}. 
\bibitem{bibq2} G.~T.~Horowitz and V.~E.~Hubeny, \pr{62} (2000) 024027;
{\tt hep-th/9909056}.
\bibitem{bibq3} G.~T.~Horowitz, \cqg{17} (2000) 1107; {\tt hep-th/9910082}. 
\bibitem{bibq4} B.~Wang, C.~Y.~Lin and E.~Abdalla, \pl{481} (2000) 79; {\tt hep-th/0003295.}
\bibitem{bibq5} B.~Wang, C.~Molina and E.~Abdalla, {\tt hep-th/0005143.}
\bibitem{bibq6} T.~R.~Govindarajan and V.~Suneeta, \cqg{18} (2001) 265; {\tt gr-qc/0007084}.
\bibitem{bibq7} V.~Cardoso and J.~P.~S.~Lemos, \pr{63} (2001) 124015;
{\tt gr-qc/0101052}.
\bibitem{bibq8} V.~Cardoso and J.~P.~S.~Lemos, \pr{64} (2001) 084017;
{\tt gr-qc/0105103}.
\bibitem{bibq9} V.~Cardoso and J.~P.~S.~Lemos, \cqg{18} (2001) 5257;
{\tt gr-qc/0107098}.
\bibitem{bibq10} J.~M.~Zhu, B.~Wang and E.~Abdalla, \pr{63} (2001) 124004;
{\tt hep-th/0101133.}
\bibitem{bibq11} B.~Wang, E.~Abdalla and R.~B.~Mann, \pr{65} (2002) 084006;
{\tt hep-th/0107243.}
\bibitem{bibq12} D.~Birmingham, \pr{64} (2001) 064024; {\tt hep-th/0101194.}
\bibitem{bibq13} D.~Birmingham, I.~Sachs and S.~N.~Solodukhin, \prl{88} (2002) 151301; {\tt hep-th/0112055}.
\bibitem{bibw1} Y.~Kurita and M.~Sakagami, \pr{67} (2003) 024003; {\tt hep-th/0208063}.
\bibitem{bibw2} D.~T.~Son and A.~O.~Starinets, {\sl JHEP \bf 0209} (2002) 042; {\tt hep-th/0205051}.
\bibitem{bibw3} V.~Cardoso, O.~J.~C.~Dias and J.~P.~S.~Lemos, \pr{67} (2003)
064026; {\tt hep-th/0212168}.
\bibitem{bibw4} V.~Cardoso, R.~Konoplya and J.~P.~S.~Lemos, {\tt gr-qc/0305037}.
\bibitem{bibq14} A.~O.~Starinets, \pr{66} (2002) 124013; {\tt hep-th/0207133}.
\bibitem{bibr1} R.~A.~Konoplya, \pr{66} (2002) 044009; {\tt hep-th/0205142.}
\bibitem{bibr2} R.~A.~Konoplya, \pr{66} (2002) 084007; {\tt gr-qc/0207028.}
\bibitem{bibus} S.~Musiri and G.~Siopsis, \pl{563} (2003) 102; {\tt hep-th/0301081.}
\bibitem{bibq15} {\em Heun's Differential Equation,} ed.~A.~Ronveaux,
Oxford University Press, Oxford (1995);\newline
S.~Y.~Slavyanov and W.~Lay, {\em Special Functions: A Unified Theory Based on Singularities,}
Oxford University Press, Oxford (2000).
\bibitem{bibmo1} L.~Motl, {\sl Adv.~Theor.~Math.~Phys.~\bf 6} (2003) 1135; {\tt gr-qc/0212096}.
\bibitem{bibmo2} L.~Motl and A.~Neitzke, {\sl Adv.~Theor.~Math.~Phys.~\bf 7} (2003) 2; {\tt hep-th/0301173}.
\bibitem{bibelse} S.~Musiri and G.~Siopsis, in preparation.
\end{thebibliography}
\end{document}